\documentclass[journal]{IEEEtran}
\usepackage{graphicx} 
\usepackage{setspace} 
\usepackage{indentfirst} 
\usepackage{listings} 
\usepackage{xcolor}  
\usepackage{courier}  
\usepackage{amsmath} 
\usepackage{svg} 
\usepackage{tabularx} 
\usepackage{makecell} 
\lstdefinelanguage{JSON}{
    basicstyle=\ttfamily\small,
    commentstyle=\color{gray},
    stringstyle=\color{red},
    numberstyle=\color{blue},
    keywordstyle=\color{orange},
    showstringspaces=false,
    breaklines=true,
    literate=
     *{0}{{{\color{blue}0}}}{1}
      {1}{{{\color{blue}1}}}{1}
      {2}{{{\color{blue}2}}}{1}
      {3}{{{\color{blue}3}}}{1}
      {4}{{{\color{blue}4}}}{1}
      {5}{{{\color{blue}5}}}{1}
      {6}{{{\color{blue}6}}}{1}
      {7}{{{\color{blue}7}}}{1}
      {8}{{{\color{blue}8}}}{1}
      {9}{{{\color{blue}9}}}{1}
      {:}{{{\color{orange}{:}}}}{1}
      {,}{{{\color{orange}{,}}}}{1}
      {\{}{{{\color{orange}{\{}}}}{1}
      {\}}{{{\color{orange}{\}}}}}{1}
      {[}{{{\color{orange}{[}}}}{1}
      {]}{{{\color{orange}{]}}}}{1},
}

\lstdefinestyle{jsonstyle}{
  language=JSON,
  basicstyle=\footnotesize\ttfamily,
  numbers=left,
  numberstyle=\tiny,
  stepnumber=1,
  numbersep=5pt,
  showstringspaces=false,
  breaklines=true,
  frame=single,
  backgroundcolor=\color{gray!10},
  literate=
   *{0}{{{\color{blue}0}}}{1}
    {1}{{{\color{blue}1}}}{1}
    {2}{{{\color{blue}2}}}{1}
    {3}{{{\color{blue}3}}}{1}
    {4}{{{\color{blue}4}}}{1}
    {5}{{{\color{blue}5}}}{1}
    {6}{{{\color{blue}6}}}{1}
    {7}{{{\color{blue}7}}}{1}
    {8}{{{\color{blue}8}}}{1}
    {9}{{{\color{blue}9}}}{1}
    {:}{{{\color{red}{:}}}}{1}
    {,}{{{\color{red}{,}}}}{1}
    {\{}{{{\color{red}{\{}}}}{1}
    {\}}{{{\color{red}{\}}}}}{1}
    {[}{{{\color{red}{[}}}}{1}
    {]}{{{\color{red}{]}}}}{1},
}

\ifCLASSINFOpdf
\else
\fi

\begin{document}

\title{LLM-Assisted Proactive Threat Intelligence for Automated Reasoning}

\author{Shuva Paul,~\IEEEmembership{Member,~IEEE,}
Farhad Alemi,~\IEEEmembership{Student~Member,~IEEE,}
and~Richard Macwan,~\IEEEmembership{Member,~IEEE}
  \thanks{Farhad Alemi is a graduate researcher at Arizona State University. \\
  Shuva Paul and Richard Macwan are researchers at the National Renewable Energy Laboratory, Golden, CO}}  
\markboth{Journal of \LaTeX\ Class Files,~Vol.~14, No.~8, August~2015}%
{Shell \MakeLowercase{\textit{et al.}}: Bare Demo of IEEEtran.cls for IEEE Journals}

\maketitle

\begin{abstract}
Successful defense against dynamically evolving cyber threats requires advanced and sophisticated techniques. This research presents a novel approach to enhance real-time cybersecurity threat detection and response by integrating large language models (LLMs) and Retrieval-Augmented Generation (RAG) systems with continuous threat intelligence feeds. Leveraging recent advancements in LLMs, specifically GPT-4o, and the innovative application of RAG techniques, our approach addresses the limitations of traditional static threat analysis by incorporating dynamic, real-time data sources. We leveraged RAG to get the latest information in real-time for threat intelligence, which is not possible in the existing GPT-4o model. We employ the Patrowl framework to automate the retrieval of diverse cybersecurity threat intelligence feeds, including Common Vulnerabilities and Exposures (CVE), Common Weakness Enumeration (CWE), Exploit Prediction Scoring System (EPSS), and Known Exploited Vulnerabilities (KEV) databases, and integrate these with the \texttt{all-mpnet-base-v2} model for high-dimensional vector embeddings, stored and queried in Milvus. We demonstrate our system's efficacy through a series of case studies, revealing significant improvements in addressing recently disclosed vulnerabilities, KEVs, and high-EPSS-score CVEs compared to the baseline GPT-4o. This work not only advances the role of LLMs in cybersecurity but also establishes a robust foundation for the development of automated intelligent cyberthreat information management systems, addressing crucial gaps in current cybersecurity practices.
\end{abstract}


\IEEEpeerreviewmaketitle

\section{Introduction}

\subsection{Context and Motivation}
In the dynamic field of cybersecurity, traditional methods often fail to address the sophisticated and evolving nature of contemporary threats \cite{idika2007survey}. Incorporating machine learning techniques, particularly large language models (LLMs), into cybersecurity practices presents a promising avenue for enhancing threat detection and response mechanisms. LLMs like BERT \cite{devlin2018bert} and GPT-3 \cite{brown2020language} have demonstrated significant advancements in comprehending and generating human-like text, which can be used to process and analyze security-related data. The development of more advanced models, such as GPT-4 \cite{achiam2023gpt}, extends these capabilities further, offering deeper contextual understanding and more effective generative functions.

The effectiveness of these models is markedly enhanced when combined with Retrieval-Augmented Generation (RAG) techniques, which merge external information with the generative capabilities of LLMs \cite{lewis2020retrieval}. We illustrate that, in cybersecurity, RAG systems can be especially beneficial in integrating real-time threat intelligence feeds, providing a more comprehensive and current understanding of the threat landscape.

Despite the advancements in LLMs and RAG systems, there is a notable gap in their application to real-time cybersecurity solutions. Traditional cybersecurity methods often struggle to keep pace with the swiftly evolving nature of cyberthreats \cite{idika2007survey}. To our knowledge, although LLMs have shown promise in static analyses, their application in dynamic, real-time threat information management remains underexplored. Current RAG systems do not fully integrate continuous threat intelligence feeds, which limits their effectiveness in real-time scenarios \cite{mitra2024localintel}.

\subsection{Research Contributions}
This research aims to develop and evaluate a novel RAG-based cyber-reasoning system that incorporates continuous threat intelligence feeds for real-time cybersecurity applications. The study seeks to bridge existing gaps in the literature by making the following contributions:
\begin{enumerate}
    \item Integrating continuous threat intelligence with RAG systems: Developing a framework that seamlessly incorporates continuous threat intelligence feeds with RAG systems to enhance real-time threat detection and response capabilities.

    \item Advancing the use of LLMs in real-time threat analysis: Extending the application of advanced LLMs, such as GPT-4o ("o" for "omni") \cite{openai2024helloGPT4o}, for continuous and dynamic threat management.

    \item Investigating the potential vulnerability of RAG-based systems against adversarial attacks and proposing mitigations.
    
\end{enumerate}

By addressing these objectives, this research aims to enhance the capabilities of automated cybersecurity systems and pave the way for future innovations in intelligent threat management.

The rest of the manuscript is organized as follows:
Section \ref{sec:background} provides an overview of fundamental concepts relevant to the study. Section \ref{sec:literature_review} examines current advancements in LLMs and RAG techniques for cybersecurity and identifies research gaps. Section \ref{sec:methodology} describes the experimental setup, data sources, and the proposed RAG-based system framework. Section \ref{sec:results} presents findings from the experiments conducted to evaluate the proposed system. Section \ref{sec:conclusion} summarizes the key findings and highlights the research contributions. Finally, Section \ref{sec:future_works} discusses potential future directions for building upon current findings and advancing the forefront of scientific research.

\section{Background} \label{sec:background}
\subsection{Cybersecurity Metrics:}
Cybersecurity standards and frameworks offer structured guidelines to safeguard information systems against cyberthreats. These frameworks enable organizations to systematically identify, manage, and mitigate risks, ensuring robust defense against increasingly sophisticated cyberattacks.

\subsubsection{Common Vulnerabilities and Exposures (CVEs)}
The CVE program was initiated by the MITRE Corporation in 1999 \cite{mitre2024, cve2024, bishop_computer_2019}. Each vulnerability under this program is assigned a unique identifier by the CVE Numbering Authority (CNA) \cite{CVEGlossaryCNA}. As of this writing, there are 255,158 published CVE entries \cite{PatrowlHearsData}.

\subsubsection{Common Weakness Enumeration (CWE)}
The CWE was developed to document the underlying weaknesses of hardware or software systems, facilitating the identification of relevant weaknesses during software development \cite{bishop_computer_2019, CWE_Weakness}. Initially, CWE started with 28 entries \cite{bishop_computer_2019} and has since expanded to 963 entries \cite{PatrowlHearsData}. CWE entries are organized hierarchically into pillar, class, base, and variant weaknesses, each level providing varying specificity to categorize and describe vulnerabilities. MITRE defines a ``weakness" as a condition that, under specific circumstances, contributes to the introduction of vulnerabilities \cite{CWE_Weakness}.

\begin{enumerate}
    \item Pillar weakness: Represents the highest-level weakness, defining an abstract theme for related weaknesses \cite{cwe_pillar_weakness}. Example: \textit{CWE-682} (Incorrect Calculation) \cite{cwe682}.
    
    \item Class weakness: Describes an issue abstractly, independent of specific technology \cite{cwe_class_weakness}. Example: \textit{CWE-922} (Insecure Storage of Sensitive Information) \cite{CWE-922}.
    
    \item Base weakness: Provides details sufficient for detecting and preventing the weakness, without specifying product type or technology \cite{cwe_base_weakness}. Example: \textit{CWE-22} (Improper Limitation of a Pathname to a Restricted Directory) \cite{CWE-22}.
    
    \item Variant weakness: Linked to a specific product type or technology, providing detailed descriptions \cite{cwe_variant_weakness}. Example: \textit{CWE-467} (Use of sizeof() on a Pointer Type) \cite{cwe-467}.
\end{enumerate}

\subsubsection{Exploit Prediction Scoring System (EPSS)}
The EPSS \cite{epss} evaluates the probability of vulnerabilities being exploited. Managed by the Forum of Incident Response and Security Teams (FIRST) \cite{firstorg}, EPSS uses statistical models and machine learning to analyze factors such as historical exploitation data and the presence of proof-of-concept exploits. Each vulnerability is assigned a score from 0 to 1, indicating the likelihood of exploitation \cite{firstorg}. EPSS data is continuously updated from sources like the National Vulnerability Database (NVD) \cite{nvd}, providing a dynamic approach to prioritizing vulnerabilities based on exploitation likelihood.

\subsubsection{Known Exploited Vulnerabilities (KEVs)}
KEVs are actively exploited by malicious actors, posing significant risks. These vulnerabilities are often found in operating systems, widely deployed applications, or network devices. Organizations like the Cybersecurity and Infrastructure Security Agency (CISA) maintain catalogs of KEVs to help prioritize security efforts \cite{cisa_website, cisa_known_exploited_vulnerabilities}. A notable example is the EternalBlue vulnerability (\textit{CVE-2017-0144}) in Microsoft's Server Message Block (SMB) protocol \cite{Nakashima_Timberg_2023, Gupta_2023, ms17-010}. Discovered by the NSA and leaked by the Shadow Brokers in 2017, it was exploited by the WannaCry ransomware attack, highlighting the severe impact of KEVs and the necessity for resilient cyber-reasoning systems \cite{Gupta_2023, liu2022working}.

\subsection{Semantics}
Embedding models and similarity metrics are essential for developing efficient cyber-reasoning systems. These models convert textual data into dense vector representations, facilitating advanced semantic analysis and retrieval.

\subsubsection{Embedding Models}
Embeddings represent words or entities as dense vectors in a continuous space, capturing syntactic and semantic properties such that similar items are mapped to near points in the vector space \cite{NIPS2000_728f206c}. Key models include Word2Vec, GloVe, and BERT. Word2Vec, developed by Mikolov et al. \cite{mikolov2013}, generates word embeddings based on contextual co-occurrences. GloVe constructs embeddings by factorizing the word co-occurrence matrix \cite{pennington2014glove}. BERT uses transformer architectures to generate contextual embeddings, capturing nuanced meanings and relationships within text \cite{devlin2018bert}.

\subsubsection{Similarity Metrics}
Effective use of embeddings requires appropriate similarity metrics. Common metrics include cosine similarity and L2 (Euclidean) distance:
\begin{itemize}
    \item Cosine similarity measures the cosine of the angle between two vectors, providing a value between -1 and 1. It is particularly useful for high-dimensional text data, focusing on orientation rather than magnitude \cite{aggarwal2001surprising}. It is defined as: 
    \[ \text{Cosine Similarity} = \frac{\mathbf{A} \cdot \mathbf{B}}{\|\mathbf{A}\| \|\mathbf{B}\|} \]
    where \( \mathbf{A} \) and \( \mathbf{B} \) are vectors, \( \mathbf{A} \cdot \mathbf{B} \) denotes the dot product, and \( \|\mathbf{A}\| \) and \( \|\mathbf{B}\|\ \) are the magnitudes of \( \mathbf{A} \) and \( \mathbf{B} \).

    \item L2 distance, also known as Euclidean distance, measures the straight-line distance between two vectors in a multidimensional space. It is effective for tasks requiring precise spatial measurements \cite{huang2008similarity}. It is calculated as:
    \[ \text{L2 Distance} = \|\mathbf{A} - \mathbf{B}\|_2 = \sqrt{\sum_{i=1}^{n} (A_i - B_i)^2} \]
    where \( A_i \) and \( B_i \) are the components of vectors \( \mathbf{A} \) and \( \mathbf{B} \).
\end{itemize}

\subsubsection{Vector Databases}
Vector databases store high-dimensional vectors, enabling efficient processing and comparison using the vector space model. Examples include Facebook AI Similarity Search (FAISS) \cite{faiss}, a popular vector database, that enables efficient similarity search through techniques like product quantization or inverted file indexing \cite{Faisspaper} for fast approximate nearest neighbor (ANN) search. Other databases like Milvus \cite{milvuspaper} and Pinecone \cite{pinecone} are designed for high performance and scalability in real-time applications, making them valuable for swiftly analyzing and retrieving critical threat intelligence entries in cyber-reasoning systems \cite{Faisspaper}.

\subsection{RAG}
RAG significantly advances natural language processing (NLP) \cite{lewis2020retrieval}. By integrating retrieval-based and generation-based methods, RAG enhances the accuracy, relevance, and richness of outputs from large language models. Traditional NLP systems either use retrieval-based methods, which are precise but limited by static data \cite{chen2017reading}, or generation-based models, which can produce plausible but incorrect information \cite{maynezfaithfulness}.

RAG models combine these approaches with a retriever and a generator. The retriever uses techniques like cosine similarity and dense passage retrieval \cite{karpukhin2020dense} to find relevant information from large external corpora. The generator then integrates this information into its outputs, improving accuracy and relevance \cite{lewis2020retrieval}.

This integration offers several benefits. RAG models efficiently handle large datasets, improving scalability and mitigating context size limitations. Grounding generation in real-world data reduces hallucinations and enhances factual accuracy \cite{lewis2020retrieval, maynezfaithfulness}. Additionally, RAG allows dynamic updates to knowledge bases without retraining, which is crucial for fields like cybersecurity, in which systems must adapt to new threats and vulnerabilities. Thus, RAG systems enable proactive and adaptable cyberdefense by leveraging advanced AI and database technologies.

\section{Related Work}
\label{sec:literature_review}
The integration of LLMs and RAG techniques in cybersecurity represents a rapidly evolving field, combining sophisticated machine learning methodologies with practical security applications. This review of the literature evaluates the leading technologies in LLM and RAG techniques for cybersecurity, examines existing frameworks, and identifies the research gaps addressed in this research.

\subsection{LLMs for Cybersecurity}
Recent advancements in LLMs have significantly influenced the cybersecurity domain. The introduction of BERT by Devlin et al. demonstrated the efficacy of bidirectional transformers in understanding and processing security-related data \cite{devlin2018bert}. BERT's deep contextual representations have been used in various cybersecurity tasks, such as threat detection and attack classification \cite{ferrag2024revolutionizing}. Building on this, GPT-3, introduced by Brown et al. \cite{brown2020language}, extended these capabilities with its substantial generative prowess, applied to tasks such as phishing detection \cite{gpt3phishing}. GPT-3's ability for few-shot learning showed that LLMs can address diverse cybersecurity challenges with minimal task-specific training data \cite{brown2020language}; few-shot learning is a technique in which a model learns to make accurate predictions with only a few training examples per class.

The development of GPT-4 \cite{achiam2023gpt} has further advanced LLM capabilities. Achiam et al. demonstrated that GPT-4 excels in handling complex and nuanced cybersecurity tasks, enhancing threat detection processes through its improved generative and contextual understanding abilities \cite{achiam2023gpt}. These advancements highlight the potential of LLMs in creating more versatile and effective cybersecurity tools.

\subsection{RAG for Cybersecurity}
RAG systems, which combine external information retrieval with generative capabilities, have emerged as a novel technique to improve LLM performance. Initial research by Lewis et al. \cite{lewis2020retrieval} showed that integrating retrieval mechanisms with generation processes significantly boosts performance across various NLP tasks by leveraging external knowledge sources. This approach has been extended to cybersecurity, where RAG systems are used to integrate threat intelligence feeds with LLMs, enhancing threat detection and response capabilities. For example, Mitra et al. \cite{mitra2024localintel} demonstrated the utility of RAG-based systems in generating threat reports by retrieving relevant information from external security databases. Although promising, the application of RAG systems in continuous threat intelligence feeds for real-time cybersecurity solutions remains underexplored.

\subsection{Continuous Threat Intelligence Feeds}
Traditional cybersecurity methods, primarily reliant on signature-based detection and heuristics, face limitations in addressing new and sophisticated threats \cite{idika2007survey}. Recent studies, such as those by Hassanin et al. \cite{hassanin2024}, investigated whether standard LLMs could be used for anomaly detection and incident response. The authors highlight that, while LLMs show significant potential, there is still room for improvement in their application for real-time scenarios.

In modern cybersecurity, threat intelligence feeds are crucial. Studies, such as those by Mitra et al. \cite{mitra2024localintel}, have examined methods for integrating threat feeds into security systems, demonstrating that updates from threat intelligence sources enhance the accuracy of threat detection and response. However, there is a pressing need for approaches that incorporate continuous threat intelligence with industrial-grade LLMs and RAG techniques to develop dynamic cybersecurity solutions.

\subsection{Gap in the Literature: RAG-Based Continuous Threat Intelligence}
Despite advancements in the application of LLMs and RAG techniques for cybersecurity, a critical gap remains: the integration of continuous threat intelligence feeds with RAG-based systems for real-time cybersecurity solutions is lacking. Existing studies primarily focus on static or batch processes rather than dynamic, continuous threat feeds \cite{mitra2024localintel}. This research proposes a RAG-based system leveraging continuous threat intelligence feeds to enhance real-time cybersecurity threat detection and response, addressing a crucial gap in current cybersecurity practices.

\section{Methodology}
\label{sec:methodology}
In this section, we address the challenges identified in the literature review by implementing a methodology designed to tackle these specific issues. Our approach aims to offer practical solutions and enhance understanding in the context of the current research.

\subsection{Threat Intelligence Retrieval}
We employed the Patrowl framework \cite{PatrowlHearsData} to automate the retrieval of cybersecurity threat intelligence feeds. This framework efficiently processes new feed additions without reprocessing the entire dataset, ensuring streamlined updates.

\subsubsection{CVE}
Our framework accesses CVE feeds from the National Vulnerability Database (NVD) \cite{nvd}, using the zipped JSON format 1.1 for its efficiency and parsing ease. The automated process systematically collects all historical entries across directories organized by year, accumulating a total of 255,158 entries \cite{PatrowlHearsData}.

\subsubsection{CWE}
CWE entries, sourced from the MITRE Corporation's \textit{cwe.mitre.org} portal \cite{cwe_website}, currently encompass 963 entries detailing both software and hardware weaknesses \cite{PatrowlHearsData}. The original data is in CSV format, with multiple columns. For our research, we focused on the ID, name, and description of each record.

\subsubsection{EPSS}
EPSS entries, curated by the Forum of Incident Response and Security Teams (FIRST) \cite{firstorg} and supported by the Cyentia Institute \cite{cyentia}, were also integrated. Typical EPSS entries include:

\begin{lstlisting}[style=jsonstyle, caption={EPSS Example}, label=EPSS-example]
"CVE-2024-26604 - EPSS": {
    "epss": "0.00042", 
    "percentile": "0.05107", 
    "date": "2024-06-28T00:00:00+0000"
},
"CVE-2004-0594 - EPSS": {
    "epss": "0.61340", 
    "percentile": "0.97823", 
    "date": "2024-06-28T00:00:00+0000"
}
\end{lstlisting}

The \textit{epss} field quantifies the exploit prediction score, with higher values indicating increased exploitability of a given vulnerability. The \textit{percentile} field provides a comparative measure for the exploitability of CVEs. The \textit{date} field marks the latest update of the EPSS. For instance, the \textit{CVE-2004-0594} vulnerability \cite{CVE-2004-0594} remains a significant threat despite its age, contrasting with newer vulnerabilities such as \textit{CVE-2024-26604} \cite{CVE-2024-26604}. This underscores the necessity of considering CVE exploitability in prioritizing mitigation strategies.\footnote{\textit{CVE-2004-0594} is a vulnerability in PHP versions up to 4.3.7 and 5.0.0RC3, permitting remote code execution when \textit{register\_globals} is enabled.} Currently, FIRST has cataloged 251,953 EPSS entries \cite{PatrowlHearsData}.

\subsubsection{KEV}
CISA \cite{cisa_website} curates a catalog of KEVs, critical for prioritizing threat mitigation. A typical KEV entry \cite{cisa_known_exploited_vulnerabilities} includes the CVE identifier, a brief description, the affected product, and the current exploitation status, such as involvement in ransomware campaigns. CISA's repository contains 1126 KEVs to date \cite{PatrowlHearsData}:
\begin{lstlisting}[style=jsonstyle, caption={KEV Example}, label=kev-example]
{
    "cveID": "CVE-2020-4428",
    "vendorProject": "IBM",
    "product": "Data Risk Manager",
    "vulnerabilityName": "IBM Data Risk Manager Remote Code Execution Vulnerability",
    "dateAdded": "2021-11-03",
    "shortDescription": "IBM Data Risk Manager contains an unspecified vulnerability which could allow a remote, authenticated attacker to execute commands on the system.",
    "requiredAction": "Apply updates per vendor instructions.",
    "dueDate": "2022-05-03",
    "knownRansomwareCampaignUse": "Unknown",
    "notes": "",
    "cwes": []
}
\end{lstlisting}

\begin{figure} 
    \centering
    \includegraphics[width=\linewidth]{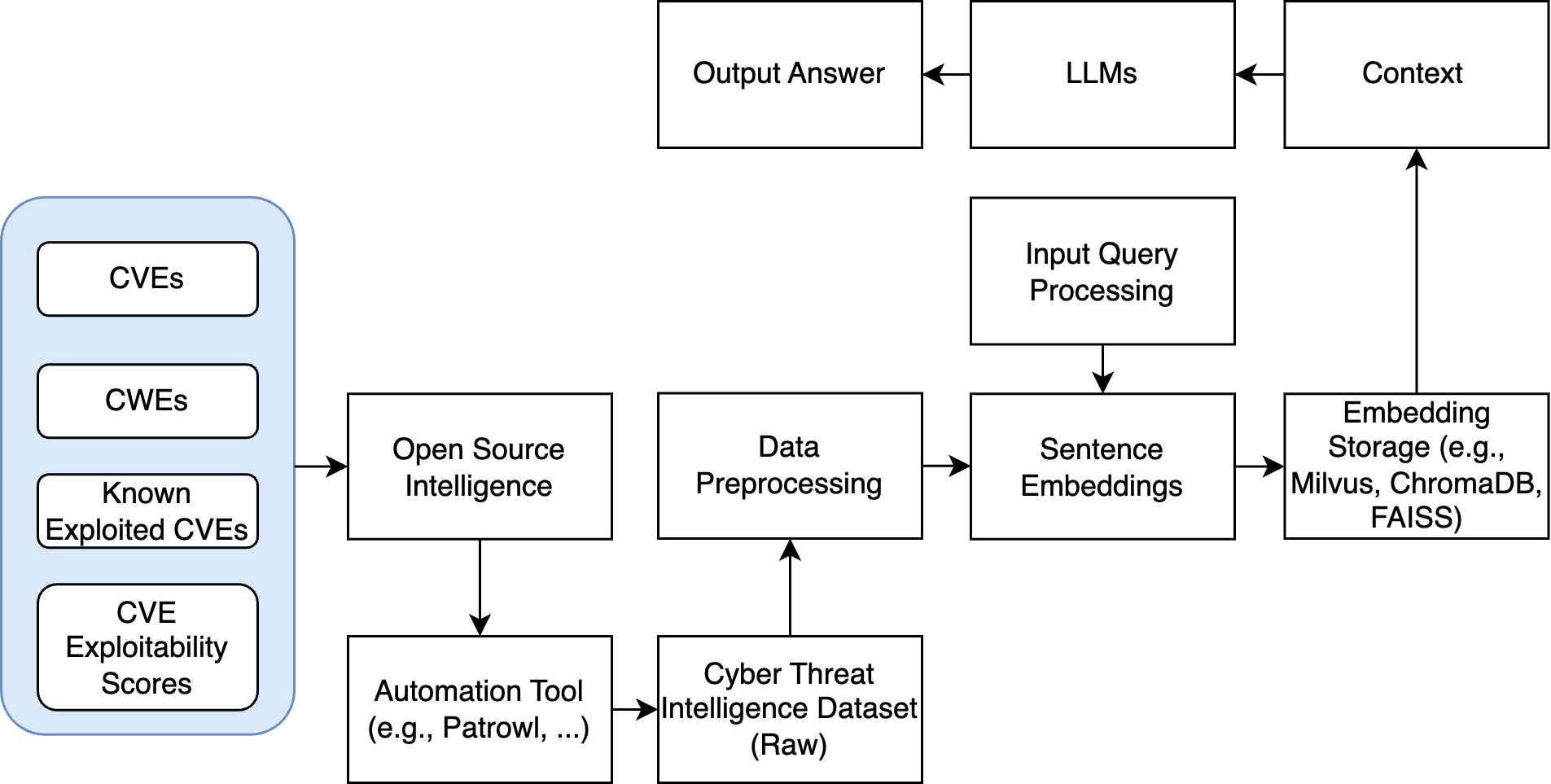}
    \caption{Threat intelligence and user query flow}
    \label{fig:method_flowchart}
\end{figure}

\subsection{Embeddings: Generation, Storage, and Retrieval}
For generating embeddings for threat intelligence feeds, the \texttt{all-mpnet-base-v2} model \cite{mpnet} from the Sentence Transformers library was employed, designed for producing high-quality sentence embeddings. This model, leveraging the MPNet architecture, generates 768-dimensional dense vector embeddings, balancing computational efficiency with semantic complexity capture \cite{mpnet, reimers_sentence}.

These embeddings are stored in Milvus, an open-source vector database optimized for high-dimensional similarity searches. Milvus scales horizontally, efficiently managing extensive vector data and performing real-time searches with high accuracy and speed. It supports various index types (e.g., IVF\_FLAT \cite{jegou2010product}, HNSW \cite{Malkov2020}, or ANNOY \cite{annoy}), balancing search precision and computational efficiency based on application needs. Additionally, Milvus's support for distributed deployments enhances performance and fault tolerance \cite{milvuspaper}. Future research is encouraged to explore distributed vector databases' efficacy in constructing large-scale cyber-reasoning systems.

User search queries are processed using the \texttt{all-mpnet-base-v2} model, converting them into vector embeddings to retain semantic meaning, as shown in Figure \ref{fig:method_flowchart}. These embeddings facilitate ANN searches in the vector database, employing cosine similarity to measure vector closeness. The vectors with the highest similarity scores are retrieved, representing the most relevant threat intelligence data. The framework ensures efficient mapping of vectors from Milvus to their original threat intelligence data points in a single database transaction. In the generation phase, these data points provide context for the user query, offering relevant information based on the semantic content.

Integrating Milvus with the \texttt{all-mpnet-base-v2} model enhances the system's ability to deliver pertinent cybersecurity threat intelligence in real-time, leveraging advanced embedding generation techniques and ensuring scalability to handle large data volumes while maintaining high retrieval performance.

\subsection{The Generation Model}
GPT-4o was selected as the generative model for the RAG system due to its proficiency in processing and generating complex domain-specific language. With a 128-k token context window \cite{openai_models}, it adeptly handles extensive contextual inputs, making it ideal for RAG-based cybersecurity threat intelligence applications.

Upon receiving a user query, the RAG system retrieves pertinent context from the Milvus database, including detailed CVE, CWE, EPSS scores, and KEV information. This additional context allows GPT-4o to generate coherent, informative, and actionable responses.

\subsection{LangChain as the Synthesizer Framework}
The LangChain framework \cite{langchain} effectively manages complex language model pipelines, facilitating seamless integration of retrieval and generation modules in RAG systems. Its modular, extensible design is crucial for scaling and optimizing retrieval and generation processes, enhancing overall system performance \cite{langchain}.

LangChain's retrieval component fetches relevant documents from an extensive corpus, integrating with the Milvus vector database to streamline the process. The generative model, GPT-4o, uses these documents to produce coherent responses. LangChain coordinates the integration of these models through its generation interface.

When a user query is received, LangChain manages preprocessing steps, including tokenization and query embedding using the \texttt{mpnet-base-v2 model}. The embedded query is then passed to the retrieval module, ensuring the retrieval of the most relevant documents. These documents are fed into the generative model to produce a coherent response, with LangChain overseeing the incorporation of document context into generative prompts. Detailed documentation of prompts used in RAG experiments and control groups using the standard GPT-4o model is provided in the appendix. Significantly, to mitigate hallucinations, prompts are designed to instruct the models to respond with "I don't know" when uncertain about an answer.
\section{Evaluations and Results}
\label{sec:results}
While many security tools perform source-level or binary-level analyses, these tools often incorrectly assume that cybersecurity researchers have access to the complete source code and the necessary infrastructure for thorough assessments. In practice, this is rarely the case due to proprietary restrictions or the high costs of comprehensive testing. Even with full source code access, source-level analysis can be impractically time-consuming. To address these challenges, we propose a practical approach that minimizes resource assumptions for cybersecurity researchers and leverages a cyber-reasoning system for intensive tasks. The following case studies illustrate our approach:

\subsubsection{Recently Disclosed CVEs}
Our cyber-reasoning system significantly outperforms the vanilla GPT-4o in addressing recently disclosed vulnerabilities. Traditional models are limited to the knowledge available at the time of their training, whereas our framework continuously integrates updated threat intelligence on zero-day vulnerabilities and emerging threats. The response generated by our system regarding \textit{CVE-2024-39471} is validated \cite{CVE-2024-39471} and documented here:
\begin{singlespace}
\begin{lstlisting}[caption={CVE-2024-39471}, label={lst:CVE-2024-39471}]
CVE-2024-39471 is a vulnerability in the Linux kernel related to the `drm/amdgpu` driver. The issue arises from an out-of-bounds read that can occur if the function `sdma_v4_0_irq_id_to_seq` returns `-EINVAL`. To mitigate this, an error handling mechanism has been added to immediately stop the process and return `-EINVAL`, thereby avoiding the out-of-bounds read.
\end{lstlisting}
\end{singlespace}

As expected, the standard GPT-4o does not have knowledge of the CVE in question due to its training data cutoff date:
\begin{singlespace}
\begin{lstlisting}[caption={Vanilla GPT-4o Response - CVE-2024-39471}, label={lst:Vanilla GPT-4o Response - CVE-2024-39471}]
I don't have information on CVE-2024-39471. It may be a new or unreleased vulnerability, or details about it might not be publicly available yet. For the latest information, you can check the official CVE database or other trusted security resources.
\end{lstlisting}
\end{singlespace}

\subsubsection{Conventional KEVs}
Regarding specific KEVs, the standard GPT-4o shows limited confidence. In contrast, our proposed framework surpasses the vanilla model by providing accurate responses. For instance, when queried about \textit{CVE-2024-38112}, which is documented as being exploited in the wild \cite{cisa_known_exploited_vulnerabilities}, our framework was able to furnish an appropriate response:
\begin{singlespace}
\begin{lstlisting}[caption={CVE-2024-38112}, label={lst:CVE-2024-38112}]
CVE-2024-38112 is a spoofing vulnerability in the Microsoft Windows MSHTML Platform. This vulnerability has a high impact on confidentiality, integrity, and availability. The recommended action is to apply mitigations as per the vendor instructions or discontinue the use of the product if mitigations are not available. The due date for these actions is July 30, 2024. There is no known ransomware campaign using this vulnerability as of now. For more details, you can refer to the [Microsoft Security Response Center](https://msrc.microsoft.com /update-guide/vulnerability /CVE-2024-38112).
\end{lstlisting}
\end{singlespace}

The vanilla model's response, however, remains basic:
\begin{singlespace}
\begin{lstlisting}[caption={Vanilla GPT-4o Response - CVE-2024-38112}, label={lst:Vanilla GPT-4o Response - CVE-2024-38112}]
I don't know.
\end{lstlisting}
\end{singlespace}

\subsubsection{CVEs With High EPSS Scores}
Vulnerabilities with high EPSS scores are of particular importance to cybersecurity professionals due to their elevated risk and the need for prioritized mitigation. We tested the new framework to identify critical vulnerabilities with high EPSS scores for proof-of-concept (PoC) purposes \cite{epss, PatrowlHearsData}. Here is the result from the proposed framework:
\begin{singlespace}
\begin{lstlisting}[caption={CVEs with High EPSS Scores}, label={lst:CVEs with High EPSS Scores}]
CVEs with high EPSS scores based on the retrieved context are:

1. CVE-2019-16758
   EPSS: 0.29708
   Percentile: 0.96961

2. CVE-2006-3581
   EPSS: 0.28304
   Percentile: 0.96888

3. CVE-2006-0671
   EPSS: 0.17811
   Percentile: 0.96193

4. CVE-2004-0599
   EPSS: 0.10541
   Percentile: 0.95061
\end{lstlisting}
\end{singlespace}

And that of the vanilla LLM:
\begin{singlespace}
\begin{lstlisting}[caption={Vanilla GPT-4o Response - CVEs with High EPSS Scores}, label={lst:Vanilla GPT-4o Response - CVEs with High EPSS Scores}]
I don't have real-time access to current data, so I can't provide a list of specific CVEs with high EPSS (Exploit Prediction Scoring System) scores at this moment. However, you can find up-to-date information on CVEs with high EPSS scores by visiting the official EPSS website or related cybersecurity resources.
\end{lstlisting}
\end{singlespace}

\subsection{Adversarial Robustness}
Adversarial robustness experiments assess how well a model handles deliberately crafted inputs designed to confuse or mislead it. We include the following robustness experiments:
\begin{itemize}
    \item Adversarial experiments using \textbf{Misleading Vulnerability Attributes} evaluate how the model handles misleading or incorrect attributes about vulnerabilities. For instance, the original prompt could be \texttt{The CVSS version 3.1 score for CVE-2024-3273 vulnerability is 9.8}. An adversarial variant could be \texttt{CVE-2024-3273 is categorized as a low-severity vulnerability.} When targeted with the adversarial variant, the proposed framework responds as follows:
    \begin{singlespace}
    \begin{lstlisting}[caption={Misleading Vulnerability Attributes}, label={lst:Misleading Vulnerability Attributes}]
    The retrieved context does not contain any information about CVE-2024-3273. Based on my knowledge, I do not have specific details regarding the categorization of CVE-2024-3273. Therefore, I cannot confirm if it is categorized as a low-severity vulnerability. If you need accurate and detailed information, I recommend checking the official CVE database or security advisories from relevant authorities.
    \end{lstlisting}
    \end{singlespace}

    As seen from the response above, the embedding model falls short of identifying the flaw in the query. As a result, non-relevant context is passed to the generator side, which in turn comes short of addressing the query. However, if we first query the framework with just the CVE entry itself (i.e., \texttt{CVE-2024-3273}) and provide that as a context for the aforementioned adversarial query, the model defends against the adversarial attack correctly:

    \begin{singlespace}
    \begin{lstlisting}[caption={Misleading Vulnerability Attributes-v2}, label={lst:Misleading Vulnerability Attributes-v2}]
    The statement that CVE-2024-3273 is categorized as a low-severity vulnerability is incorrect. According to the provided context, CVE-2024-3273 is a critical vulnerability with a CVSS v3.1 base score of 9.8.
    \end{lstlisting}
    \end{singlespace}

    This experiment indicates that the embedding component has a room for improvement and can be made more robust with a more powerful embedding model.

    \item Adversarial experiments can also purposefully include \textbf{Ambiguous Queries} to test the framework's handling of vague queries pertaining to vulnerabilities. For instance, a vague query could be \texttt{Tell me about CVE-2024-3273's resolution.} Here is what the proposed framework has to say:
    \begin{singlespace}
    \begin{lstlisting}[caption={Ambiguous Queries}, label={lst:Ambiguous Queries}]
    The retrieved context does not provide any information about CVE-2024-3273. Based on my own knowledge and the absence of details in the provided context, I cannot provide information on CVE-2024-3273's resolution. If you need specific details, I recommend checking the official CVE database or resources provided by the software vendor associated with the CVE.
    \end{lstlisting}
    \end{singlespace}

    As can be seen above, the generator complains from a lack of proper context. This is because the vague query confuses the embedding model and makes it miss the (already-present) context in the vector space. However, when we pass the correct context to the model using a chain-of-thought process, the model correctly understands the vague query:
    \begin{singlespace}
    \begin{lstlisting}[caption={Ambiguous Queries-v2}, label={lst:Ambiguous Queries-v2}]
    CVE-2024-3273 is a critical vulnerability in certain D-Link NAS devices (DNS-320L, DNS-325, DNS-327L, and DNS-340L). The resolution for this vulnerability, as indicated in the retrieved context, is as follows:

    The affected products are no longer supported by the vendor and have reached their end-of-life (EOL) or end-of-service (EOS) lifecycle. The vendor has confirmed that these products should be retired and replaced. Therefore, the required action for mitigating this vulnerability is to retire and replace the affected devices according to the vendor's instructions.

    For more detailed information, you can refer to the support announcement provided by D-Link: [D-Link Support Announcement](https://support announcement.us.dlink.com/security /publication.aspx?name=SAP10383).

    \end{lstlisting}
    \end{singlespace}

\end{itemize}

\subsection{Further Discussion}
A key advantage of our proposed framework is its complete modularity, allowing for internal maintenance, augmentation, and replacement of individual components without affecting the overall system:
\begin{itemize}
    \item Additional cybersecurity intelligence sources such as Feedly \cite{Feedly_Threat_Intelligence} and FireEye Threat Intelligence (now Trellix) \cite{trellix_threat_intelligence_services} can be seamlessly integrated.
    
    \item Any embedding model that is compatible with the Sentence Transformers library API \cite{reimers-2019-sentence-bert} can be used, including proprietary models such as \texttt{text-embedding-3} or \texttt{text-embedding-ada-002} from OpenAI \cite{openai_models}.
    
    \item The vector database Milvus can be deployed locally, within a Docker container, in the cloud, or substituted with a different database \cite{milvuspaper}.
    
    \item The generator model is flexible, supporting locally running open-source LLMs (e.g., \texttt{Llama-3 70B}) \cite{meta_llama_3_70b}, models from the OpenAI family \cite{openai_models}, or custom cloud deployments such as Azure \cite{microsoft2024ai}.

    \item A human user can be substituted with an LLM agent, facilitating integration into a multi-agent system for fully autonomous cyber-reasoning using frameworks like Autogen \cite{wu2023autogen}.

    \item The framework supports deployment as a scheduled task or cron-job for real-time threat intelligence.
\end{itemize}

These capabilities are enabled by the modular design and the automation of the RAG mechanism via LangChain.
\section{Conclusion}
\label{sec:conclusion}
This research elucidates the promising integration of LLMs with RAG systems within the cybersecurity domain. Our empirical findings indicate that coupling LLMs with RAG systems and continuous threat intelligence feeds can substantially enhance real-time threat detection and response. This integration allows for a transition from static threat analysis to a dynamic, real-time threat management paradigm, fostering a more proactive and adaptive security stance.

Furthermore, the coupling of RAG systems with continuous threat intelligence feeds, initially proposed by Mitra et al. \cite{mitra2024localintel} and extended in our study, offers a robust framework for using extensive external data to support automated cybersecurity decisions. This methodology addresses the shortcomings of traditional signature-based and heuristic approaches, which often falter against the swiftly evolving cyberthreat landscape \cite{idika2007survey}.

Our work bridges significant gaps in current cybersecurity solutions by introducing an innovative RAG-based system that leverages continuous threat intelligence feeds for real-time application. This not only advances the role of LLMs in cybersecurity but also paves the way for future advancements in automated intelligent threat management systems. We urge the broader scientific community to build upon our findings, refine the proposed framework, and investigate advanced-prompt engineering techniques to achieve superior outcomes.

\section{Future Works}
\label{sec:future_works}
This paper explores the feasibility of cyber-reasoning systems driven by threat intelligence feeds through experimental validation. Further research could focus on evaluating these systems' effectiveness with tools like binary analysis frameworks and fuzzers, as well as their proficiency in navigating directories and URLs.

Future work could extend this research into a multi-agent paradigm, in which a user proxy agent communicates tasks to a managerial LLM. The LLM would divide tasks among specialized worker agents with skills in filesystem navigation, tool use, and RAG usage, ensuring high parallelizability and resilience to agent failures. The Microsoft Autogen framework \cite{wu2023autogen} offers a starting point for developing such a multi-agent system.

Additionally, fine-tuning LLMs for cybersecurity and using those LLMs to generate RAG prompts shows promise. Moreover, future studies could leverage advanced embedding models from the HuggingFace Massive Text Embedding Benchmark Leaderboard \cite{huggingface_mteb_leaderboard} for more sophisticated embedding generation. Finally, the current prompts can be improved using advanced prompt engineering techniques for achieving superior results.
\section*{Acknowledgments}
This work was authored by the National Renewable Energy Laboratory, operated by Alliance for Sustainable Energy, LLC, for the U.S. Department of Energy (DOE) under Contract No. DE-AC36-08GO28308. This work was supported by the Laboratory Directed Research and Development Program at NREL. The views expressed in the article do not necessarily represent the views of the DOE or the U.S. Government. The U.S. Government retains and the publisher, by accepting the article for publication, acknowledges that the U.S. Government retains a nonexclusive, paid-up, irrevocable, worldwide license to publish or reproduce the published form of this work, or allow others to do so, for U.S. Government purposes.


\clearpage
\onecolumn
\appendix

\subsection*{Appendix A}
\subsubsection*{Proposed Framework Query-Prompt}
System Prompt
\begin{lstlisting}[style=jsonstyle, caption={Proposed Framework Query-Prompt}, label={lst:Proposed Framework Query-Prompt}]
"""You are an assistant specializing in question-answering tasks. Use the provided context to answer the question. If the context does not contain the answer, rely on your own knowledge. If you are still NOT very sure of your answer, just say "I don't know"; DO NOT GUESS. 

Question: {question} 
Context: {context} 
Answer:
"""
\end{lstlisting}

\subsubsection*{Vanilla GPT-4o Query-Prompt}
System Prompt
\begin{lstlisting}[style=jsonstyle, caption={Vanilla GPT-4o Query-Prompt}, label={lst:Vanilla GPT-4o Query-Prompt}]
"""You are an assistant specializing in question-answering tasks. Use your own knowledge to answer the question. If you are NOT very sure of your answer, just say "I don't know"; DO NOT GUESS.

# Question: {question}
# Answer:
"""
\end{lstlisting}

\end{document}